\documentclass[lettersize,journal]{IEEEtran}
\usepackage{amsmath,amsfonts,amssymb}
\usepackage{algorithm}
\usepackage{algpseudocode}
\usepackage{array}
\usepackage{caption}
\usepackage{float}
\usepackage{stfloats}
\usepackage{textcomp}
\usepackage{cuted}
\usepackage{stfloats}
\usepackage{url}
\usepackage{enumerate}
\usepackage{verbatim}
\usepackage{graphicx}
\usepackage{cite}
\usepackage{textcomp}
\usepackage{xcolor}
\usepackage{bm}
\usepackage{subcaption}

\usepackage{subfloat}
\usepackage{balance}
\usepackage{url}
\usepackage{caption}

\usepackage{cite}
\usepackage {epstopdf}
\usepackage{orcidlink} %
\hypersetup{hidelinks} % hide the hyperlink box
\usepackage{color}
\usepackage{balance}
\usepackage{xcolor}
\allowdisplaybreaks[4]

\usepackage{etoolbox}

\makeatother


\ifCLASSINFOpdf

\else

\fi

\begin{document}
	\title{\huge Movable Antenna Enhanced Dual-Functional  \\ Radar-Communication: A Symbol-Level Precoding Approach}
	\author{Ran Yang,  Ning Wei, \IEEEmembership{Member,~IEEE}, Zheng Dong, \IEEEmembership{Member,~IEEE}, \\Chadi Assi, \textit{Fellow, IEEE}, You Li, Fei Xu, and Yue Xiu, \IEEEmembership{Member,~IEEE} \vspace{-16pt}
		\thanks{Ran Yang, Yue Xiu, and Ning Wei are with the National Key Laboratory of Wireless Communications, University of Electronic Science and Technology of China, Chengdu 611731, China (e-mail: yangran6710@outlook.com; wn@uestc.edu.cn; xiuyue12345678@163.com). 
			
		Zheng Dong is with the School of Information Science and Engineering, Shandong University, Qingdao 266237, China (e-mail: zhengdong@sdu.edu.cn). Chadi Assi is with Concordia University, Montreal, Quebec, H3G 1M8, Canada (e-mail: assi@ciise.concordia.ca). Fei Xu is with ZGC Institute of Ubiquitous-X Innovation and Applications, Beijing 100081, China (e-mail: xufei@zgc-xnet.com). You Li is with Southwest China Research Institute of Electronic Equipment (SWIEE), Chengdu, China.}}
	
	\markboth{Journal of \LaTeX\ Class Files,~Vol.~18, No.~9, September~2020}%
	{How to Use the IEEEtran \LaTeX \ Templates}
	\maketitle
	\vspace{-2mm}	
\begin{abstract}
This letter investigates a symbol-level {precoder} design for movable antenna (MA)-enhanced dual-functional radar-communication (DFRC) {systems}. To enhance radar sensing capabilities, we formulate an optimization problem aimed at maximizing the minimum radar signal-to-interference-plus-noise ratio (SINR) across multiple targets in a cluttered environment. Our approach jointly designs the space-time {transmitted} waveforms, receiving filters, and antenna placement. However, the resulting problem is intractable {to solve} due to practical waveform constraints and the non-linear mapping from {antenna positions to the corresponding channel coefficients.} To address these challenges, we develop a bi-level optimization framework {by} leveraging deep reinforcement learning (DRL). Specifically, the twin delayed deep deterministic policy gradient (TD3) algorithm is employed in the outer layer to optimize antenna placement, while penalty convex-concave procedure (CCP) and majorization-minimization (MM) techniques are incorporated in the inner layer for {regularizing} waveform design. Simulation results demonstrate that the proposed method significantly improves radar SINR and achieves a superior sensing-communication trade-off compared to benchmark schemes.
\end{abstract}
	
	\begin{IEEEkeywords}
		Dual-functional radar-communication,  symbol-level precoding, movable antenna.
		
	\end{IEEEkeywords}
	
	\vspace{-13.0pt}	
	\section{Introduction}
	\IEEEPARstart{A}{CCORDING} to the International Mobile Telecommunications (IMT)-2030 vision, future sixth-generation (6G) wireless networks are envisioned to support intelligent radio environments with integrated communication, localization, and perception capabilities~\cite{8869705}. In particular, dual-functional radar–communication (DFRC), which unifies sensing and communication within a single hardware platform, has emerged as a fundamental enabler for 6G systems. By sharing spectrum resources and signal-processing modules, DFRC systems have demonstrated significant potential in diverse applications such as vehicle-to-everything (V2X), remote healthcare, and intelligent transportation systems~\cite{10636212}.
	
	However, most existing DFRC approaches rely on block-level precoding (BLP), wherein waveforms are designed based on second-order {signal} statistics~\cite{9585321}. While BLP is effective for conventional communication systems that transmit random symbols, radar sensing typically demands deterministic probing sequences with favorable temporal characteristics. Consequently, BLP-based schemes inherently suffer from degraded sensing performance. Moreover, the linear processing  in BLP provides  limited spatial {degrees-of-freedom (DoFs)} and multiplexing gains, thereby constraining {the} overall system performance. To address this issue, symbol-level precoding (SLP) techniques~\cite{9035662}, capable of exploiting all available DoFs in both spatial and temporal domains, have been employed to boost the dual-task performance~\cite{9769997,10902060,10844869}. Despite their demonstrated advantages, existing SLP-based designs predominantly assume conventional fixed-position antennas (FPAs), which restrict adaptability to dynamic sensing and communication environments~\cite{10643473}. Additionally, the static array geometry fails to mitigate spatial conflicts between radar and communication directions, leading to simultaneous degradation in array gain and quality-of-service (QoS).
	
	Recently, movable antennas (MAs) {have been introduced} to overcome the limitations of FPAs~\cite{10286328}. In a {typical MA-assisted } system, each antenna element is connected to a radio frequency (RF) chain via flexible cables, enabling active repositioning in real time~\cite{10354003}. A hardware prototype of an MA-enabled system was demonstrated in~\cite{11224420}, and corresponding channel modeling and performance analysis were investigated in~\cite{10318061}. Building upon these foundations, several studies have explored MA-enhanced DFRC systems (see~\cite{11373884,11249714,2026arXiv260114868Y,11418623,11353414,10870338} and references therein). Nevertheless, these works {are still} confined to BLP-based designs. To the best of our knowledge, SLP-based waveform design for MA-assisted DFRC systems remains largely unexplored, motivating the development of a tailored MA-enhanced symbol-level signaling scheme to unlock the full potential of integration between sensing and communication.
	
	 In this letter, we investigate symbol-level signal design for an MA-enhanced DFRC system. Our main contributions are summarized as follows: 1) We formulate a joint optimization problem to maximize the minimum radar signal-to-interference-plus-noise ratio (SINR) across multiple targets in a cluttered environment, subject to communication QoS and peak-to-average power ratio (PAPR) constraints. The optimization is achieved through the joint design of transmitted waveforms, receiving filters, and antenna placement. 2) To tackle the intractable problem, we develop a bi-level structured deep reinforcement learning (DRL)-based framework, where the outer layer employs the twin delayed deep deterministic policy gradient (TD3) algorithm to optimize antenna placement and the inner layer leverages penalty-based convex–concave procedure (CCP) and majorization–minimization (MM) techniques for waveform design. 3) Numerical results demonstrate that the proposed method significantly improves radar SINR and achieves a superior trade-off between sensing and communication performance compared to benchmark schemes.

	\section{System Model}
	{We consider} a movable antenna-enhanced DFRC system, where the base station (BS)  is equipped with two separate MA-based linear arrays dedicated to signal transmission and reception, respectively. Each array is composed of $N$ MAs. The dual-functional BS serves $K$  users while simultaneously detecting $W$ point-like targets in the presence of $Q$ clutters. The feasible movement range for MAs is a one-dimensional (1D) interval of length $D$. The transceiver antenna positioning vectors (APVs) are denoted by $\boldsymbol{t} = [t_1,t_2,\dots,t_N]^T \in \mathbb{R}^{N\times1}$ and $\boldsymbol{r} = [r_1,r_2,\dots,r_N]^T \in \mathbb{R}^{N\times1}$, respectively, with $0\le t_1 \le t_2\le \dots \le t_N \le D$ and $0\le r_1 \le r_2\le  \dots \le  r_N \le D$.
	\subsection{Communication Model}
	Given that the signal propagation distance is significantly larger than the size of moving regions, the far-field response is adopted for channel modeling\cite{10318061}. Specifically, the angle-of-arrival (AoA), angle-of-departure (AoD), and amplitude of the complex coefficient for each link remain constant despite antenna movement. The geometric model in \cite{10354003} is employed for communication channels, where the number of  scattering paths at transceiver nodes is the same. Denote by $L_k$ the number of propagation paths between the BS and user $k$, where the azimuth angle of the $j$-th path at the  BS is given by $\psi_{k}^j\in [0,\pi]$.  Then, the signal propagation difference between the position of the $n$-th transmitting MA $t_n$ and the reference point $o^t$ is given by $		\rho(t_n,\psi_k^j) = t_n\cos\psi_{k}^j,\forall k,j,n.$ Consequently, the field response vector (FRV) at $t_n$ can be given by 
	\begin{equation}
		\boldsymbol{g}_k(t_n) = \left[e^{\jmath\frac{2\pi}{\lambda}\rho(t_n,\psi_k^1)},\dots,e^{\jmath\frac{2\pi}{\lambda}\rho(t_n,\psi_k^{L_k})}\right]^T\in \mathbb{C}^{L_k\times 1},
	\end{equation} 
	where $\lambda$ is the carrier wavelength. Therefore,  the field response matrix (FRM) of the link from the BS to user $k$ for all $N$ transmitting MAs is given by
	\begin{equation}
		\boldsymbol{G}_k({\boldsymbol{t}})\triangleq\left[\boldsymbol{g}_k({t}_1),\boldsymbol{g}_k({t}_2),\dots,\boldsymbol{g}_k({t}_{N})\right]\in \mathbb{C}^{L_k\times N}.
	\end{equation}
	Let $\boldsymbol{\Sigma}_k = \text{diag}\{\sigma_{k,1},\sigma_{k,2},\dots,\sigma_{k,L_k}\}\in\mathbb{C}^{L_k\times L_k}$ denote the path response matrix (PRM), and the channel matrix between the BS and the $k$-th user is given by 
	\begin{equation}
		\boldsymbol{h}_k^T({\boldsymbol{t}}) = \boldsymbol{1}_{L_k}^T\boldsymbol{\Sigma}_k\boldsymbol{G}_k({\boldsymbol{t}})\in \mathbb{C}^{1\times N}, 1\le k \le K,\label{communicationchannel}
	\end{equation}
	where the all-ones vector $\boldsymbol{1}_{L_k}\in\mathbb{R}^{L_k\times 1}$ characterizes the FRV associated with the $L_k$ scattering paths to the $k$-th user. 
	Define that the discrete waveform $\boldsymbol{X} \triangleq \big[\boldsymbol{x}[1],\dots,\boldsymbol{x}[M]\big]\in \mathbb{C}^{N\times M}$, where  $M$ represents the length of the transmission block. Thus, the received signal at the $k$-th user in the $m$-th time slot is given by 
	\begin{equation}
		r_k[m] = \boldsymbol{h}_k^T(\boldsymbol{t})\boldsymbol{x}[m]+n_k[m],1\le m \le M,\label{received_signal}
	\end{equation}
	where $n_k[m]\sim\mathcal{CN}(0,\sigma_k^2)$ is the additive white Gaussian noise (AWGN). We assume that the communication data is independently selected from an $\Omega$-phase-shift-keying (PSK) constellation. For the $K$ communication users, the symbols transmitted at the $m$-th time slot are denoted by $\boldsymbol{s}[m] = \big[s_1[m],s_2[m],\dots,s_K[m]\big]^T\in\mathbb{C}^{K\times1}$. In particular, the SLP method is employed to convert multi-user interference (MUI) into  constructive interference (CI) that enhances communication QoS~\cite{9035662}. The CI principle can be characterized by 
	\begin{align}\label{CI1}
		&\Re \left\lbrace \bm {h}_{k}^{T}(\boldsymbol{t}) \bm {x}[m]e^{-\jmath \angle {s_{k}[m]}} {\kern-1.42271pt}-\sigma_k\sqrt{\gamma _{k}}\right\rbrace \sin \Phi \notag\\ 
		\, &~~-\big |\Im \left\lbrace \bm {h}_{k}^{T}(\boldsymbol{t}) \bm {x}[m]e^{-\jmath \angle {s_{k}[m]}}\right\rbrace \big |{\kern-1.42271pt}\cos \Phi \geq 0, \, \forall k,m, 
	\end{align} 
	where $\gamma_k$ is the required QoS of the $k$-th user and $\Phi = \pi/\Omega$. The constraints in~\eqref{CI1}  {can be equivalently} rewritten as 
	\begin{equation} \label{CI2}
		\Re \left\lbrace \boldsymbol {h}_{k}^{T}(\boldsymbol{t}) \bm {x}[m]\frac{e^{-\jmath \angle s_{k}[m]} (\sin \Phi\pm e^{-\jmath \pi /2}\cos \Phi)}{\sigma_k\sqrt{\gamma _{k}}\sin \Phi }\right\rbrace \!\geq\! 1, 
	\end{equation} 
	In order to explicitly express the constraint in~\eqref{CI2}, we first ignore the AWGN $n_k[m]$ in~\eqref{received_signal}, and the received noise-free signal $\tilde{r}_k[m]$ at the $m$-th time slot can be given by
	\begin{equation}
		\tilde{r}_k[m] = [\boldsymbol{e}_m^T\otimes\boldsymbol {h}_{k}^{T}(\boldsymbol{t})] \boldsymbol{x},\forall k,m,
	\end{equation}
	where $\boldsymbol{x} = \operatorname{vec}(\boldsymbol{X})\in\mathbb{C}^{NM\times 1}$, and $\boldsymbol{e}_m$ is the $m$-th column of an $M$-dimensional identity matrix $\bm{I}_M$. Let $\boldsymbol{h}_{k,m}(\boldsymbol{t}) = \boldsymbol{e}_m\otimes\boldsymbol {h}_{k}(\boldsymbol{t})$ and $\gamma_{k,m} = \frac{e^{-\jmath \angle s_{k}[m]} (\sin \Phi\pm e^{-\jmath \pi /2}\cos \Phi)}{\sigma_k\sqrt{\gamma _{k}}\sin \Phi }$, and the CI constraints in~\eqref{CI2} can be recast as 
	\begin{equation}
		\Re \left\lbrace \gamma_{k,m} \boldsymbol{h}_{k,m}^T(\boldsymbol{t})\boldsymbol{x}\right\rbrace \!\geq\! 1,\forall k, m.
	\end{equation}
	\vspace{-9mm}
	\subsection{Radar Model} Since each radar pulse has $M$ digital samples, the range domain can be divided into $M$ discrete bins. The target $w$ and  clutter $q$ are  located in the range-angle positions $(\kappa_w,\theta_w)$ and $(v_q,\xi_q)$, respectively, where $1\le w\le W$, $1\le q \le Q$, and we assume that $\kappa_1\le \kappa_2\le \ldots \le \kappa_W$. We adopt the line-of-sight (LoS) channel model for sensing links between the BS and targets/clutters\cite{10318061}. Then, the receiving and transmitting steering vectors are respectively given by $\boldsymbol{a}_r(\omega,\boldsymbol{r}) = [e^{\jmath\frac{2\pi}{\lambda}\rho(r_1,\omega)},\dots,e^{\jmath\frac{2\pi}{\lambda}\rho(r_N,\omega)}]^T\in\mathbb{C}^{N\times 1}$ and $\boldsymbol{a}_t(\omega,\boldsymbol{t}) = [e^{\jmath\frac{2\pi}{\lambda}\rho(t_1,\omega)},\dots,e^{\jmath\frac{2\pi}{\lambda}\rho(t_N,\omega)}]^T\in\mathbb{C}^{N\times 1}$, where $\omega$ is the azimuth angle for the target/clutter. As such, the channels for the $w$-th target return and the  $q$-th clutter return  can be respectively given by 
	\begin{align} 
		\mathcal {F}_{w}(\boldsymbol {t,r })\triangleq&~\boldsymbol{a}_r(\theta_w,\boldsymbol{r})\boldsymbol{a}_t(\theta_w,\boldsymbol{t})^T,1\le w \le W,\label{target_channel}\\ 
		\mathcal {C}_{q}(\boldsymbol {t,r })\triangleq&~\boldsymbol{a}_r(\xi_q,\boldsymbol{r})\boldsymbol{a}_t(\xi_q,\boldsymbol{t})^T,1 \le q \le Q.\label{clutter_channel}
	\end{align}
	
	The echo signal observed at the BS is given by
	\begin{small}
		\begin{equation}
			\mathbf{Y} = \sum_{w=1}^{W}\alpha_w\mathcal{F}_w(\boldsymbol{t,r})\tilde{\mathbf{X}}\mathbf{J}_{\kappa_w-\kappa_1} + \sum_{q=1}^{Q}\alpha_q \mathcal{C}_q(\boldsymbol{t,r})\tilde{\mathbf{X}}\mathbf{J}_{v_q-\kappa_1} +\mathbf{N}_R.\notag
		\end{equation} 
	\end{small}Note that: \textit{a)} $\alpha_w$ and $\alpha_q$ denote  the complex reflection coefficients for the target $w$ and the clutter $q$, respectively, which capture both the round-trip path loss and corresponding radar cross section (RCS).  {We assume that $\mathbb{E}\{|\alpha_w|^2\} = \zeta_w^2$ and $ \mathbb{E}\{|\alpha_q|^2\} = \varsigma_q^2$}; \textit{b)} $\tilde{\mathbf{X}} = [\mathbf{X},~\mathbf{0}_{N\times(\kappa_W-\kappa_1)}]\in \mathbb{C}^{N\times\tilde{M}}$ denotes the zero-padded waveform matrix, where $\tilde{M} = M+\kappa_W-\kappa_1$; \textit{c)} $\mathbf{J}_{\kappa_w-\kappa_1}\in \mathbb{C}^{\tilde{M}\times\tilde{M}}$ is the shift matrix whose $(i,j)$-th element is defined by ${\mathbf{J}}_{\kappa_w-\kappa_1}(i,j) \triangleq \left\{ \begin{array}{ll} 1,& \text{if } j - i = \kappa_w-\kappa_1,\\ 0,& {\text{{otherwise.}}} \end{array} \right.$; \textit{d)} $\mathbf{N}_R = [\mathbf{n}_R[1],\dots,\mathbf{n}_R[\tilde{M}]]\in \mathbb{C}^{N\times\tilde{M}}$ is the AWGN with $\mathbf{n}_R[m]\sim\mathcal{CN}(\bm{0}_{N\times1},\sigma_r^2\mathbf{I}_{N})$.
	
	%	\begin{itemize}
		%		\item $\alpha_t$ and $\alpha_q$ denote  the complex reflection amplitudes of the target $t$ and the clutter $q$, respectively. {For simplicity, in this paper, we assume that $\mathbb{E}\{|\alpha_t|^2\} = \mathbb{E}\{|\alpha_q|^2\} = \zeta^2$.}
		%		\item  $\tilde{\mathbf{X}} = [\mathbf{X},~\mathbf{0}_{N\times(r_T-r_1)}]\in \mathbb{C}^{N\times\tilde{L}}$ denotes the zero-padded Tx waveform matrix, where $\tilde{L} = L+r_T-r_1$.	
		%		\item $\mathbf{J}_{r_t-r_1}\in \mathbb{C}^{\tilde{L}\times\tilde{L}}$ is the shift matrix whose $(i,j)$-th element is defined by
		%		\begin{equation*} 
			%			{\mathbf{J}}_{r_t-r_1}(i,j) = \left\{ \begin{array}{ll} 1,& \text{if } j - i = r_t-r_1,\\ 0,& {\text{{otherwise.}}} \end{array} \right. 
			%		\end{equation*}
		%		\item $\mathbf{N}_R = [\mathbf{n}_R[1],\dots,\mathbf{n}_R[\tilde{L}]]\in \mathbb{C}^{N\times\tilde{L}}$ is the additive white Gaussian noise (AWGN) with $\mathbf{n}_R[l]\sim\mathcal{CN}(0,\sigma_r^2\mathbf{I}_{N})$, 1$\le l\le\tilde{L}$, where $\mathbf{I}_N$ is the $N$-dimensional identity matrix. 
		%	\end{itemize}
	Stacking the received echo signals by defining $\boldsymbol{y} \triangleq \rm{vec}(\mathbf{Y})$, $\bm{n} \triangleq \rm{vec}(\mathbf{N}_R)$, and $\mathbf{T} \triangleq [\mathbf{I}_{NM},\mathbf{0}_{NM\times N(\kappa_W-\kappa_1)}]^T\in\mathbb{C}^{N\tilde{M}\times NM}$, the received signals  can be recast as
	\begin{equation} 
		\boldsymbol{y} = \sum _{w=1}^{W}\alpha _{w}\widetilde {\mathcal {F}}_{w}(\boldsymbol {t,r })\boldsymbol {x} + \sum _{q=1}^{Q}\alpha _{q}\widetilde {\mathcal {C}}_{q}(\boldsymbol {t,r })\boldsymbol {x} + \boldsymbol {n},
	\end{equation} 	
	where $\widetilde {\mathcal {F}}_{w}(\boldsymbol {t,r }) = (\mathbf {J}_{\kappa_1-\kappa_w} \otimes \mathcal {F}_{w}(\boldsymbol {t,r }))\mathbf{T}$ and $\widetilde{\mathcal {C}}_{q}(\boldsymbol {t,r }) = (\mathbf {J}_{\kappa_1-v_q} \otimes \mathcal {C}_{q}(\boldsymbol {t,r }))\mathbf{T}$.
	In order to detect the target $w$, a linear space-time receiving filter $\boldsymbol{u}_w \in \mathbb{C}^{N\tilde{M}\times 1}$ is applied to process the received signal. As such, the associated radar output is $\boldsymbol{u}_w^H\boldsymbol{y}$, given by
	\begin{equation*}
		\boldsymbol{u}_w^H\boldsymbol{y} = \boldsymbol{u}_w^H\sum _{w=1}^{W}\alpha _{w}\widetilde {\mathcal {F}}_{w}(\boldsymbol {t,r })\boldsymbol {x} + \boldsymbol{u}_w^H\sum _{q=1}^{Q}\alpha _{q}\widetilde {\mathcal {C}}_{q}(\boldsymbol {t,r })\boldsymbol {x} + \boldsymbol{u}_w^H\boldsymbol {n},
	\end{equation*}
	and the SINR $\Gamma_{w}(\bm{x},\bm{t},\bm{r},\bm{u}_w)$ is shown in (\ref{radar sinr output}) at the top of this page.
	\begin{figure*}[t]
		\begin{equation}\label{radar sinr output}
			\Gamma_{w}(\bm{x},\bm{t},\bm{r},\bm{u}_w) = \frac{\zeta_w^2|\boldsymbol{u}_w^H\widetilde {\mathcal {F}}_{w}(\bm{t,r})\bm{x}|^2}
			{\bm{u}_w^H[\sum _{i=1,i\neq w}^{W}\zeta_i^2\widetilde{\mathcal {F}}_{i}(\bm{t,r})\bm{x}\bm{x}^H\widetilde{\mathcal {F}}_{i}(\bm{t,r})^H+\sum_{q=1}^{Q}\varsigma_q^2\widetilde {\mathcal {C}}_{q}(\boldsymbol {t,r })\bm{x}\bm{x}^H\widetilde {\mathcal {C}}_{q}(\boldsymbol {t,r })^H+\sigma_r^2\mathbf{I}_{N\tilde{M}}]\bm{u}_w}.
		\end{equation}
		\hrule
		\vspace{-0.5cm}
	\end{figure*}

	\subsection{Problem Formulation}
	%	\vspace{-1mm}
	Based on the above performance metrics,  the transmission waveform $\bm{x}$, transceiver APVs $\bm{t}$ and $\bm{r}$, and the receiving filter $\bm{u} \triangleq [\bm{u}_1^T,\dots,\bm{u}_W^T]^T$ are jointly designed to maximize the minimum radar SINR while ensuring the PAPR  and  QoS constraints. The optimization problem is formulated as 
	\begin{subequations}\label{Problem1}
		\begin{alignat}{2}
			&\underset{\bm {x},\bm{u}, \bm {t,r }}{\max} ~~\underset{w}{\min} ~~\Gamma _{w}(\bm{x},\bm{t},\bm{r},\bm{u}_w)\label{P1_objective}\\
			&\,\text {s.t.}~ ~\Re \left\lbrace \gamma_{k,m} \boldsymbol{h}_{k,m}^T(\boldsymbol{t})\boldsymbol{x}\right\rbrace \!\geq\! 1, \forall m,k, \label{P1_CI}\\
			&\hphantom {s.t.~}
			\bm{x}^H\bm{x} = P_tM,\label{P1_PAPR1}\\
			&\hphantom {s.t.~}
			|x_i| \le \sqrt{\frac{P_t\eta}{N}}, 1\le i \le NM,\label{P1_PAPR2}\\
			&\hphantom {s.t.~} t_1\ge 0,t_N\le D, r_1\ge 0, r_N\le D,\label{P1_MA1}\\
			&\hphantom {s.t.~} t_n-t_{n-1}\ge d, r_n-r_{n-1}\ge d, 2 \le n \le N,\label{P1_MA2}
		\end{alignat} 
	\end{subequations} 
	where $P_t$ is the available transmission power, $\eta$ is the customized parameter for controlling the PAPR level, and $d$ represents the minimum distance between MAs to prevent coupling effect. The problem in (\ref{Problem1}) is highly intractable due to the non-concavity of the objective function in (\ref{P1_objective}),  equality restriction in (\ref{P1_PAPR1}), as well as non-linear mapping from antenna placement to channel coefficients.
	\section{Bi-level DRL-based Algorithm}
	To solve the intractable problem in~\eqref{Problem1}, we first transform the problem in~\eqref{Problem1} into a more favourable form. Then, we develop a bi-level structured DRL-based algorithm, the details of which are elaborated as follows.
	\vspace{-2mm}
	\subsection{Problem Reformulation}  
	For the problem in~\eqref{Problem1}, we first note that the receiving filters $\{\boldsymbol{u}_w\}_{w=1}^W$ are independent of each other and only exist in the objective function in~\eqref{P1_objective}. Therefore, the {optimization} on $\{\boldsymbol{u}_w\}_{w=1}^W$ can be formulated as a minimum variance distortionless response (MVDR) problem\cite{11373884}. The optimal solution $\bm{u}_w^\star$ can be directly  given by (\ref{filter}), {shown at the bottom of this page.}
	\begin{figure*}[hb]
		\hrule
		\vspace{0.1cm}
		\begin{small}
			\begin{equation}\label{filter}
				\boldsymbol{u}_{w}^{\star} = \frac{[\sum _{i=1,i\neq w}^{W}\zeta_i^2\widetilde{\mathcal {F}}_{i}(\bm{t},\bm{r})\bm{x}\bm{x}^H\widetilde{\mathcal {F}}_{i}(\bm{t,r})^H+\sum_{q=1}^{Q}\varsigma_q^2\widetilde {\mathcal {C}}_{q}(\boldsymbol {t,r})\bm{x}\bm{x}^H\widetilde {\mathcal {C}}_{q}(\boldsymbol {t,r })^H+\sigma_r^2\mathbf{I}_{N\tilde{M}}]^{-1}{\widetilde{\mathcal {F}}}_{w}(\bm{t,r})\bm{x}}
				{\bm{x}^H\widetilde{\mathcal{F}}_w(\boldsymbol{t,r})^H[\sum _{i=1,i\neq w}^{W}\zeta_i^2\widetilde{\mathcal {F}}_{i}(\bm{t,r})\bm{x}\bm{x}^H\widetilde{\mathcal {F}}_{i}(\bm{t,r})^H+\sum_{q=1}^{Q}\varsigma_q^2\widetilde {\mathcal {C}}_{q}(\boldsymbol {t,r})\bm{x}\bm{x}^H\widetilde {\mathcal {C}}_{q}(\boldsymbol {t,r })^H+\sigma_r^2\mathbf{I}_{N\tilde{M}}]^{-1}\widetilde{\mathcal{F}}_w(\boldsymbol{t,r})\boldsymbol{x}}.
			\end{equation}
		\end{small}
		\hrule
		\vspace{0.05cm}
		\begin{small}
			\begin{equation}\label{f}
				f_w(\bm{x},\boldsymbol{t,r}) = \zeta_w^2 \bm{x}^H\widetilde{\mathcal{F}}_w{(\boldsymbol{t,r})}^H{\underbrace{[\sum _{i=1,i\neq w}^{W}\zeta_i^2\widetilde{\mathcal {F}}_{i}(\bm{t,r})\bm{x}\bm{x}^H\widetilde{\mathcal {F}}_{i}(\bm{t,r})^H+\sum_{q=1}^{Q}\varsigma_q^2\widetilde {\mathcal {C}}_{q}(\boldsymbol {t,r })\bm{x}\bm{x}^H\widetilde {\mathcal {C}}_{q}(\boldsymbol {t,r })^H+\sigma_r^2\mathbf{I}_{N\tilde{M}}]}_{\triangleq\bm{R}_w}}{}^{-1} \underbrace{\widetilde{\mathcal{F}}_{w}(\boldsymbol{t,r})\bm{x}}_{\triangleq\bm{z}_w}.
			\end{equation}
		\end{small}
	\end{figure*}
	By substituting  $\bm{u}_w^{\star}$ into the objective function in~\eqref{P1_objective}, denoted by $f_w(\bm{x},\bm{t,r})$ in (\ref{f}), the problem in~\eqref{Problem1} can be recast as
	\begin{subequations}\label{Problem2}
		\begin{alignat}{2}
			&\underset{\bm {x}, \boldsymbol {t,r }}{\max} ~~\underset{w}{\min} ~~f_w(\bm{x},\boldsymbol{t,r})\label{P2_obj}\\
			&\,\text {s.t.}~\eqref{P1_CI},~\eqref{P1_PAPR1},~\eqref{P1_PAPR2},~\eqref{P1_MA1},~\eqref{P1_MA2}.
		\end{alignat} 
	\end{subequations} 
	
	Unfortunately, the problem in~\eqref{Problem2} is still difficult to solve due to the coupling of $\bm{x}$, $\bm{t}$, and $\bm{r}$. Compared with conventional BLP schemes, the SLP design involves many hard affine constraints and much higher optimization dimensionality. Therefore, directly putting all optimization variables into a DRL framework faces unstable convergence issues, and may cause dimensionality curse. 
	To deal with this issue, we propose a bi-level DRL optimization framework. Specifically, in the outer level, the APV design is reformulated as a Markov decision process (MDP), and the TD3 framework is adopted to determine the antenna positions. In the inner level, a penalty CCP-based MM method is developed to optimize transmit waveform $\bm{x}$, thereby accelerating training convergence.
	\subsection{TD3-based Antenna Placement}
	The transceiver antenna placement problem is reformulated as an MDP. The state space, action space, reward function, and neural-network training procedure are specified as follows.
	\subsubsection{Action Space $\mathcal{A}$} In each step, the TD3 agent adjusts the action defined by
	\begin{equation}
		\bm{a}_{\tau} = \{\bm{t}_{\tau},\bm{r}_{\tau}\},
	\end{equation}
	where  $\bm{t}_{\tau}$ and $\bm{r}_{\tau}$ are the transmit and receive antenna position action at step $\tau$, respectively.
	\subsubsection{State Space $\mathcal{S}$} The state space provides the agent with essential information for  policy selection in a concise form
	\begin{equation}
		\bm{s}_{\tau} = \{\Re\{\bm{\sigma}\},\Im\{\bm{\sigma}\},\bm{\vartheta},\bm{a}_{\tau-1}\},
	\end{equation}
	where $\bm{\sigma}$ and $\bm{\vartheta}$ denote the complex  path-loss  coefficients and azimuth angles, respectively. Specifically, they are defined as
	$\bm{\sigma} = [\bm{\sigma}_1^T,\dots,\bm{\sigma}_K^T,\{\zeta_w^2\}_{w=1}^W,\{\varsigma_q^2\}_{q=1}^Q]^T$,
	$\bm{\sigma}_k = [\sigma_{k,1},\dots,\sigma_{k,L_k}]^T$,
	$\bm{\vartheta} = [\bm{\psi}_1^T,\dots,\bm{\psi}_K^T,\{\theta_w\}_{w=1}^W,\{\xi_q\}_{q=1}^Q]^T$, and
	$\bm{\psi}_k = [\psi_k^1,\dots,\psi_k^{L_k}]^T$.
	
	\subsubsection{Reward $\mathcal{R}$}  The reward provides feedback from the environment to guide the agent policy. The reward function is constructed as
	\begin{equation}
		R_{\tau} = \min~\{ f_w(\bm{x},\boldsymbol{t,r}) \}_{w=1}^{W},
	\end{equation} 
	which can be obtained once the transmit waveform is determined via \textit{Algorithm 2} in  Section III-C.  Note that the  constraints in~\eqref{Problem2} are inherently fulfilled in different levels of the algorithm.  Specifically, for~\eqref{P1_MA1} and~\eqref{P1_MA2}, 
	the transceiver APVs are constructed through a group of displacement variables $\{\Delta_n^{(s)}\}_{n=1}^N$  as follows
	\begin{equation}
		s_n =  (n-1)d + \sum_{k=1}^{n}\Delta_k^{(s)}, 1 \le n \le N,
	\end{equation}
	where $s \in \{t,r\}$. As such, the constraints in~\eqref{P1_MA1} and~\eqref{P1_MA2} can be equivalently written  as $\{\Delta_n^{(s)}\}_{n=1}^{N} \ge 0$ and $\sum_{n=1}^{N}\Delta_n^{(s)} \le \Delta_{\max}$ with $\Delta_{\max} = D - (N-1)d$, which can be implemented by a sigmoid-softmax transformation.   The constraints in~\eqref{P1_CI},~\eqref{P1_PAPR1}, and~\eqref{P1_PAPR2} are enforced within the penalty CCP framework.
	\subsubsection{Training Mechanism}
	A TD3-based framework is employed for agent training. Let $Q_{i}$ and $Q_{i}'$, $i=1,2$, 
	represent the two critic networks and their corresponding target critic 
	networks, respectively, where $c_i$ and $c_i'$ denote the associated 
	network parameters. During training procedure, $c_i$ is updated by minimizing 
	the mean squared Bellman error, i.e., 
	$\mathcal{L}(c_i)
	=\mathbb{E}_{(\bm{s},\bm{a},R,\bm{s}')\sim \mathcal{D}}
	\left[\left(Q_{i}(\bm{s},\bm{a})-y
	\right)^2\right],$ where $(\bm{s},\bm{a},R,\bm{s}')$ denotes $\mathcal{B}$ transitions sampled from the replay buffer $\mathcal{D}$, and 
	$y = R + \rho' \min_{i=1,2} Q_{i}^\prime(\bm{s}^\prime,\tilde{\bm{a}}^{\prime})$ is the target Q-value.
	Here, $\rho'$ is the discount factor, and $\tilde{\bm{a}}^{\prime}$ denotes the smoothed target action obtained from the target actor policy $\pi_{\phi^{\prime}}$, i.e., 
	\begin{equation}
		\tilde{\bm{a}}^{\prime} = \pi_{\phi^{\prime}}(\bm{s}^{\prime}) + \varepsilon,\label{target_action}
	\end{equation}
	where $\varepsilon$ is the clipped Gaussian noise with $\varepsilon\sim\text{clip}(\mathcal{N}(0,\sigma_{cn}^2),-c_{cn},c_{cn})$. The actor policy $\pi_{\phi}$ is optimized to maximize the Q-value predicted by $Q_1$, i.e.,
	\begin{equation} 
		\max _{{\phi} }{\,}{\,}\mathbb {E}_{\bm{s}\sim \mathcal {D}}\left [{{Q_{{1}}\left ({{\bm{s},\pi _{\phi }(\bm{s})}}\right)}}\right ]. \label{actor_update}
	\end{equation}
	Note that the actor and all target networks are updated at a reduced frequency compared  to critic networks, with a delay factor $d_p$. In particular, the target networks are updated via
	\begin{equation}
		c_{i}^{\prime} \leftarrow \iota c_{i} + (1-\iota)c_{i}^{\prime}, \phi^{\prime}  \leftarrow  \iota \phi + (1-\iota)\phi^{\prime}, i = 1,2,\label{target_update}
	\end{equation}
	where $ \iota$ is the soft update coefficient. To maintain effective exploration after the nonlinear action mapping, the Ornstein-Uhlenbeck (OU) noise is applied before  logits are mapped into concrete actions, i.e., 
	$\sigma _{ou}\left ({{\varrho  }}\right) = \sigma _{ou,\text{ini}} e^{-\varpi \varrho } + \sigma _{ou,\min },$
	where $\varrho $ is the episode index, $\varpi$ is the decay rate, and $\sigma _{ou,\text{ini}}$ and $\sigma _{ou,\min }$ are the initial and minimum noise power, respectively. The training procedure is summarized in \textit{Algorithm 1}.
	\begin{algorithm}[t] \small
		\caption{DRL-Based Antenna Placement Optimization}
		\begin{algorithmic}[1]
			\State \textbf{Initialize:} Replay buffer $\mathcal{D}$, actor network $\pi_{\phi}$, critic networks $Q_{1}, Q_{2}$, and target networks $\pi_{\phi^{\prime}}, Q_{1}^{\prime}, Q_{2}^{\prime}$.
			\For{each episode $\varrho = 0,1,\ldots,N_{epi}-1$}
			\State Initialize antenna positions $\bm{t}_0$ and $\bm{r}_0$, and set exploration noise power according to $\sigma _{ou}\left ({{\varrho  }}\right)$.
			\For{each time step $\tau = 0,1,\dots,N_{ts}-1$}
			\State Generate $\bm{a}_{\tau}$ according to $\pi_{\phi}(\bm{s}_\tau)$ and $\sigma_{ou}(\varrho)$;
			\State  Execute $\bm{a}_{\tau}$ and observe $\bm{s}_{\tau+1}$;
			\State Obtain the  transmit waveform $\bm{x}$ via \textit{Algorithm 2};
			\State Calculate $R_{\tau}$, and  store $(\bm{s}_{\tau},\bm{a}_{\tau},R_{\tau},\bm{s}_{\tau+1})$ into $\mathcal{D}$;
			\State Sample mini-batch $(\bm{s},\bm{a},R,\bm{s}^{\prime})$ from $\mathcal{D}$;
			\State Build target actions with clipped noise via~\eqref{target_action};
			\State Compute $y$ 
			and update $c_i$ via $\mathcal{L}(c_i)$;
			\If{$\tau \bmod d_p = 0$}
			\State Update the actor network via~\eqref{actor_update};
			\State Update target networks via~\eqref{target_update};
			\EndIf
			\EndFor
			\EndFor
			\State \textbf{Output:} Actor policy $\pi_{\phi}$ for antenna placement.
		\end{algorithmic}
	\end{algorithm}
	\subsection{Transmission Waveform Optimization}
	With APVs being fixed, the problem in~\eqref{Problem2} with respect to  $\bm{x}$ can be formulated as 
	\begin{subequations}\label{Problem3}
		\begin{alignat}{2}
			&\underset{\bm {x},\alpha}{\max} ~~ \alpha \label{P3_OBJ}\\
			&\,\text {s.t.}~f_w(\bm{x}) \ge \alpha,\forall w,\label{P3_SINR} \\
			&\hphantom {s.t.~}
			~\eqref{P1_CI},~\eqref{P1_PAPR1},~\eqref{P1_PAPR2},
		\end{alignat} 
	\end{subequations}
	where $\alpha$ is an auxiliary variable. We note that the non-convexity of the problem in~\eqref{Problem3} lies in~\eqref{P3_SINR} and~\eqref{P1_PAPR1}. To deal with this issue, a penalty CCP method is proposed. Specifically, we first construct a  global lower bound for  $f_w(\bm{x})$ by utilizing the following lemma.
	
	\textit{\textbf{Lemma 1}}: For a positive-definite matrix $\boldsymbol{R}$, $\bm{z}^H\bm{R}^{-1}\bm{z}$ is a convex function of $\bm{z}$ and $\bm{R}$, and its surrogate function at the point $(\bm{z}^{(p)},\bm{R}^{(p)})$ is given by
	\begin{small}
		\begin{align}\label{Lemma1}
			\bm {z}^{H}\bm {R}^{-1}\bm {z}&  \geq 2\Re \left\lbrace  (\bm{z}^{(p)})^{H}(\bm {R}^{(p)})^{-1}\bm {z}\right\rbrace\notag\\ &-\text{Tr}\left\lbrace (\bm{R}^{(p)})^{-1} \bm {z}^{(p)}(\bm {z}^{(p)})^{H}(\bm {R}^{(p)})^{-1}\bm {R}\right\rbrace.
		\end{align} 
	\end{small}
	
	Combining~\eqref{Lemma1} and~\eqref{f}, the surrogate objective function $\tilde{f}_w(\bm{x})$ can be given by
	\begin{equation*}
		{f}_w(\bm{x})\ge  \tilde{f}_w(\bm{x}) \triangleq -\bm{x}^H\bm{G}_w^{(p)}\bm{x} + \Re\{(\bm{b}_w^{(p)})^H\bm{x}\} +  c_w^{(p)},
	\end{equation*}
	where $\bm{b}_w^{(p)} = 2\zeta_w^2\tilde{\mathcal{F}}_w(\bm{t},\bm{r})^H(\bm{R}_w^{(p)})^{-1}\bm{z}_w^{(p)}$, $c_w^{(p)} = -\zeta_w^2\sigma_r^2(\bm{z}_w^{(p)})^H(\bm{R}_w^{(p)})^{-1}(\bm{R}_w^{(p)})^{-1}\bm{z}_w^{(p)}$, $\bm{z}_w^{(p)} = \tilde{\mathcal{F}}_w(\bm{t},\bm{r})\bm{x}^{(p)}$, $\bm{R}_w^{(p)}$ is the value of $\bm{R}_w\in\mathbb{C}^{N\tilde{M}\times N\tilde{M}}$ at the point $\bm{x}^{(p)}$, and $\bm{G}_w^{(p)}\in\mathbb{C}^{NM\times NM}$ is shown in~\eqref{Gwp} at the top of next page.
	\begin{figure*}[tp]
		\begin{small}
			\begin{equation}	\label{Gwp}
				\bm{G}_{w}^{(p)}= \sum_{i=1,i\ne w}^{W}\zeta_w^2\zeta_i^2\widetilde{\mathcal{F}}_i(\bm{t},\bm{r})^H(\bm{R}_{w}^{(p)})^{-1}\bm{z}_{w}^{(p)}(\bm{z}_{w}^{(p)})^H(\bm{R}_{w}^{(p)})^{-1}\widetilde{\mathcal{F}}_i(\boldsymbol{t},\bm{r})+
				\sum_{q=1}^{Q}\zeta_w^2\varsigma_q^2\widetilde{\mathcal{C}}_q(\boldsymbol{t},\bm{r})^H(\bm{R}_{w}^{(p)})^{-1}\bm{z}_{w}^{(p)}(\bm{z}_{w}^{(p)})^H(\bm{R}_{w}^{(p)})^{-1}\widetilde{\mathcal{C}}_q(\boldsymbol{t},\bm{r}).
			\end{equation}
		\end{small}
		\hrule
	\end{figure*}
	
	Next, we move on to deal with the equality restriction in~\eqref{P1_PAPR1}. Specifically, we split~\eqref{P1_PAPR1} into two constraints, i.e., $\bm{x}^H\bm{x} \le P_tM$ and $\bm{x}^H\bm{x} \ge P_tM$. The latter reverse-convex constraint is  linearized  by using the first-order Taylor expansion as
	$\Re\{2(\bm{x}^{(p)})^H\bm{x}-(\bm{x}^{(p)})^H\bm{x}^{(p)}\} \ge P_tM.$	
	Following the penalty CCP framework\cite{boyd2004convex}, we impose the use of a slack variable $\beta\ge 0$ over the equivalent
	constraints of equality restriction, which yields
	\begin{subequations}\label{Problem4}
		\begin{alignat}{2}
			&\underset{\bm {x},\alpha,\beta}{\max} ~~ \alpha - \mu^{(p)} \beta \label{P4_OBJ}\\
			&\,\text {s.t.}~\tilde{f}_w(\bm{x}) \ge \alpha,\forall w,\label{P4_SINR} \\
			&\hphantom {s.t.~}
			\bm{x}^H\bm{x}\le P_tM + \beta,\label{P4_PAPR1}\\
			&\hphantom {s.t.~}
			\Re\{2(\bm{x}^{(p)})^H\bm{x}-(\bm{x}^{(p)})^H\bm{x}^{(p)}\} \ge P_tM - \beta,\\
			&\hphantom {s.t.~}
			~\eqref{P1_CI},~\eqref{P1_PAPR2},~\beta \ge 0.
		\end{alignat} 
	\end{subequations}
	Here, $\mu^{(p)}$ is the regularization factor to scale the impact of the penalty term $\beta$, which controls the feasibility of the constraints. The problem in~\eqref{Problem4} is convex and can be solved by off-the-shelf solvers~\cite{boyd2004convex}. Note that: \textit{a)} An upper bound limit $\mu_{\text{max}}$ is imposed to avoid numerical problems, that is, a feasible solution may not be found if $\mu^{(p)}$ grows too large; \textit{b)} The penalty CCP belongs to MM with $\mu^{(p)} = \mu_{\text{max}}$, and the  obtained solution is guaranteed to  converge to a  stationary point of the problem in~\eqref{Problem3}\cite{boyd2004convex}; \textit{c)} The computational complexity is given by $\mathcal{O}(\sqrt{2W+M(2N+K)}(M^3N^3W+M^3N^2K))$.

	\begin{algorithm}[t] \small
		\caption{Penalty CCP Procedure for Transmit Waveform}
		\begin{algorithmic}[1]
			\State \textbf{Initialize:} set $p = 0$, $\delta > 1$, and initialize $\boldsymbol{x}^{(p)}$.
			\Repeat
			\State Update $\bm{x}^{(p+1)}$, $\alpha$, and $\beta$ from Problem in~\eqref{Problem4};
			\State $\mu^{(p+1)} = \min\{\delta\mu^{(p)},\mu_{\max}\}$;
			\State $p = p + 1$;
			\Until{$\beta \leq \chi$ and $\Vert\bm{x}^{(p)} - \bm{x}^{(p-1)}\Vert \leq \upsilon$}.
			\State Calculate $\bm{u}_w^\star, \forall w$, by \eqref{filter}.
			\State \Return $\boldsymbol{x}^\star$ and $\bm{u}^\star$.
		\end{algorithmic}
	\end{algorithm}
	\vspace{-2mm}
	\section{Numerical Results}
	In this section, simulation results are carried out to evaluate the performance of the proposed design.  We compare our scheme with three baseline schemes: \textbf{1) SLP-FPA}: With FPAs spaced between intervals of $\lambda/2$, \textit{Algorithm 2} is directly employed to optimize transmission waveform $\bm{x}$; \textbf{2) BLP-MA}: The BLP and transceiver antenna positions are jointly optimized to enhance the radar performance~\cite{11373884}; \textbf{3) BLP-FPA}: The BLP method is performed with FPAs to maximize the radar SINR performance. 
	\captionsetup{font={small},labelsep=period}
	\captionsetup[subfloat]{font=small}
	\begin{figure}
		\centering 
		\subfloat[]{ %
			\begin{minipage}[t]{0.45\linewidth}%
				\centering \includegraphics[width=1.5in]{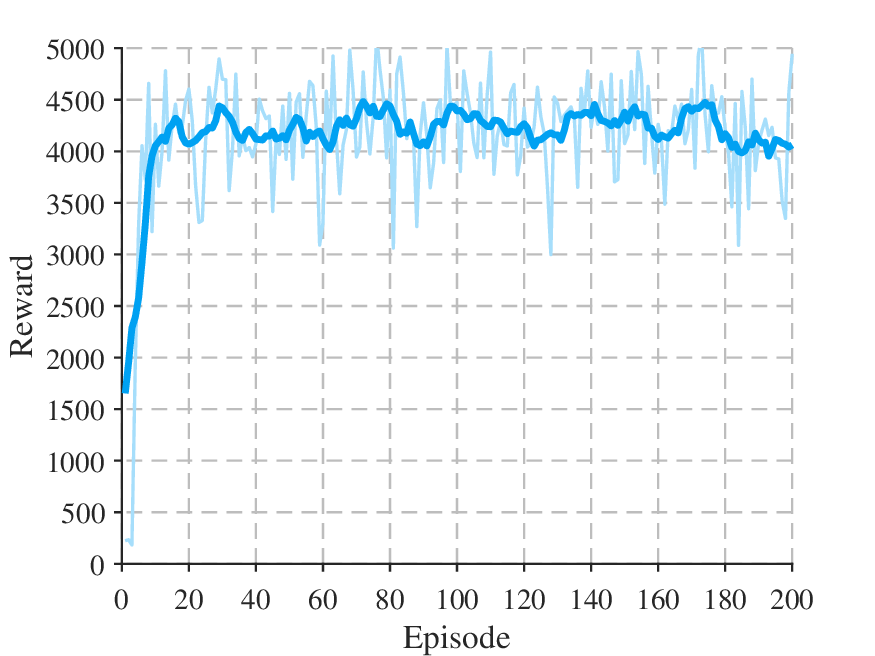} \label{convergence}
		\end{minipage}}
		\subfloat[]{ %
			\begin{minipage}[t]{0.45\linewidth}%
				\centering \includegraphics[width=1.5in]{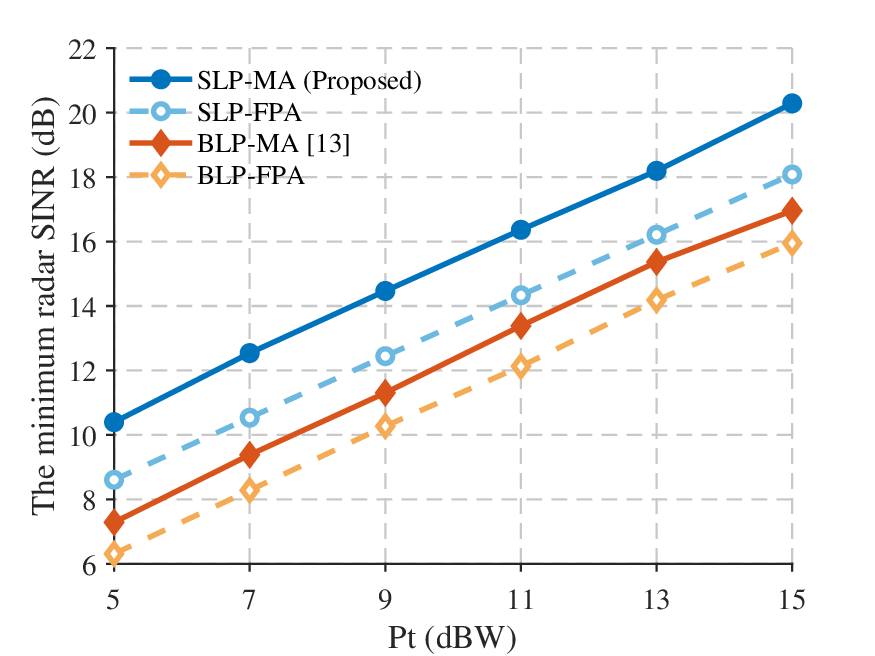} \label{Pt}
		\end{minipage}}
		\caption{(a) Convergence behaviour of the proposed algorithm. (b) The minimum radar SINR versus transmission power $P_t$.}
		\label{SSS1}
	\end{figure}
	\vspace{-1mm}
	\begin{figure}
		\centering 
		\subfloat[]{ %
			\begin{minipage}[t]{0.45\linewidth}%
				\centering \includegraphics[width=1.5in]{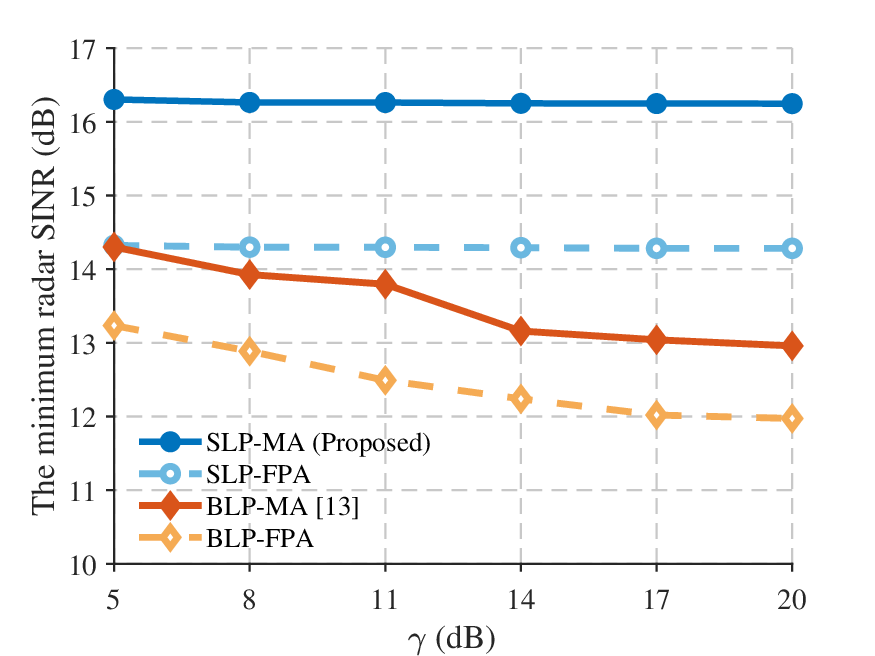}\label{gamma}
		\end{minipage}}
		\subfloat[]{ %
			\begin{minipage}[t]{0.45\linewidth}%
				\centering 
				\includegraphics[width=1.5in]{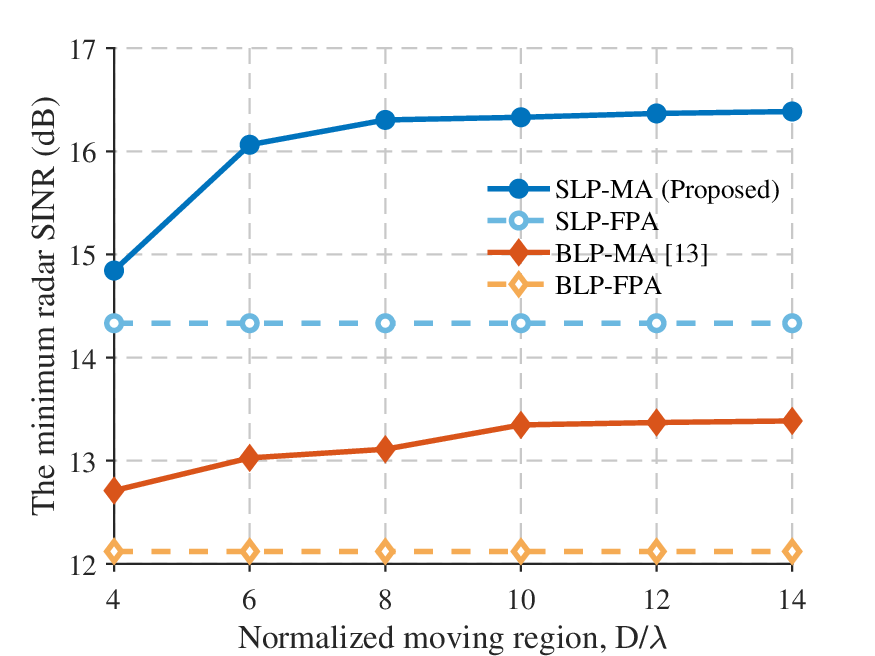} \label{region}
		\end{minipage}}
		\caption{(a) The minimum radar SINR versus QoS $\gamma$. (b) The minimum radar SINR versus normalized moving region.}
		\label{SSS2}
	\end{figure}
	
	In our simulation, the path loss model is given by  $PL = C_0d_{link}^{-\nu}$, where $C_0 = -30$ dB and the path loss exponents for radar and communication links are given by 2.6 and 3.2, respectively. The BS is located at $(0, 0)$ m. The users are randomly distributed in a circle centered at $(40,0)$ m with a radius of 5 m. Other parameters are used unless specified: $N = 6, M = 10, W = 2, Q = 3, K = 5, L = 8, \Omega = 4, d_p = 3,  \iota = 0.002,N_{epi} = 200, N_{ts} = 100, \lambda = 0.1\text{m}, D = 12\lambda, \eta = 2.2, \sigma_r^2 = \sigma_k^2 =  -90~\text{dBm},\text{and}~\gamma_k =  15~\text{dB},\forall k.$ The range-angle positions of targets and  clutters are given by $\{(0,70^{^\circ}),(0,100^{^\circ})\}$ and $\{(1,105^{^\circ}), (0,30^{^\circ}),(1,75^{^\circ})\}$ respectively. The learning rate  are set as $2\times10^{-5}$ and $1.5\times10^{-4}$ for actor and critic networks, respectively. All networks take 2 hidden layer with dimension 768 activated by ReLU.

	We first present the convergence behaviour of the proposed algorithm in Fig.\,\ref{SSS1}(\subref{convergence}). It can be observed that the training 
	reward stabilizes after around 30 episodes, corresponding to  3000
	environment interactions. This fast stabilization benefits from the bi-level architecture of the proposed algorithm, 
	which substantially reduces the dimensionality of the action space. Fig.\,\ref{SSS1}(\subref{Pt}) plots  the minimum radar SINR  versus $P_t$. It can be observed that the proposed SLP-MA scheme significantly outperforms the other benchmark schemes. 
	This is attributed to the active movement of antennas, which not only enhances the desired channel gains, but also provides spatial DoFs to mitigate interference. 
	
	We plot the radar performance versus QoS $\gamma$ in Fig.\,\ref{SSS2}(\subref{gamma}). Compared with BLP-based schemes, the proposed SLP scheme achieves a better  
	radar-communication trade-off,  which highlights the   superiority of the proposed method. Fig.\,\ref{SSS2}(\subref{region}) demonstrates the minimum radar SINR versus the normalized moving region. The result shows that the optimal performance can be attained  within finite regions, indicating   that a moderately sized moving region can be selected to strike a satisfactory balance between performance and costs.
	\vspace{-3mm}
	%	\begin{figure}[!t]
		%		\centering
		%		\includegraphics[width=0.45\textwidth]{convergence.eps}
		%		\captionsetup{font={normalsize},labelsep=period,singlelinecheck=off}
		%		\caption{Training reward convergence of the proposed algorithm.}
		%		\label{convergence} 
		%	\end{figure}%
	%	
	%	\begin{figure}[!t]
		%		\centering
		%		\includegraphics[width=0.45\textwidth]{Pt.eps}
		%		\captionsetup{font={normalsize},labelsep=period,singlelinecheck=off}
		%		\caption{Training reward convergence of the proposed algorithm.}
		%		\label{Pt} 
		%	\end{figure}%
	%		\begin{figure}[!t]
		%		\centering
		%		\includegraphics[width=0.45\textwidth]{gamma.eps}
		%		\captionsetup{font={normalsize},labelsep=period,singlelinecheck=off}
		%		\caption{Training reward convergence of the proposed algorithm.}
		%		\label{Gamma} 
		%	\end{figure}%
	%	
	%		\begin{figure}[!t]
		%		\centering
		%		\includegraphics[width=0.45\textwidth]{region.eps}
		%		\captionsetup{font={normalsize},labelsep=period,singlelinecheck=off}
		%		\caption{Training reward convergence of the proposed algorithm.}
		%		\label{Region} 
		%	\end{figure}%

	%\begin{figure}[htbp]
	%	\centerline{\includegraphics[width=2.5in]{RIS.eps}}
	%	\captionsetup{font={normalsize},labelsep=period}
	%	\caption{RIS} 
	%	\label{element}
	%\end{figure}
	%	\vspace{-2mm}
	\section{Conclusion}
	In this letter, we have proposed a unified framework for designing joint transmit waveforms, receiving filters, and antenna placement to maximize the minimum radar SINR subject to PAPR and QoS constraints. A bi-level structured DRL-based method was developed to solve the challenging problem.  Simulation results demonstrated that the proposed method can significantly improve the radar SINR, and achieve a superior trade-off between sensing and communication performance. 
	\vspace{-3.5mm}
	\balance
	\bibliography{reference} 
	\bibliographystyle{IEEEtran} 
	
	% \bibliography{reference}
	\vfill
\end{document}